\documentclass[a4paper]{jpconf}
\usepackage{amstext}

\usepackage{amssymb,epsfig,graphicx}
\usepackage{amsmath}
\usepackage{amscd}
\usepackage{verbatim}
\usepackage{lscape}

        \newcommand{\hsp}{\hspace{.2 in}}
        
        \newcommand{\beq}{\begin{eqnarray}}
        \newcommand{\eeq}{\end{eqnarray}}
        
         \newcommand{\cs}{ coherent states }

        \newcommand{\ket}[1]{$ |{#1}\rangle$}
        
        \newcommand{\mbra}[1]{\langle {#1}|}
        \newcommand{\mket}[1]{ |{#1}\rangle}

        \newcommand{\sks}[2]{\langle {#1}|{#2}\rangle }
        
        \newcommand{\op}[1]{$\hat{#1}$}
        \newcommand{\mop}[1]{\hat{#1}}

\begin{document}

\title{Coherent states on the circle}


\author{G Chadzitaskos, P Luft and J Tolar}

\address{Department of Physics\\
Faculty of Nuclear Sciences and Physical Engineering         \\
Czech Technical University in Prague\\ B\v rehov\'a 7,  CZ - 115 19
Prague, Czech Republic}

\ead{jiri.tolar@fjfi.cvut.cz, goce.chadzitaskos@fjfi.cvut.cz}

\begin{abstract}

We present a possible construction of \cs on the unit circle as
configuration space. In our approach the phase space is the product
$\mathbb{Z} \times S^{1}$. Because of the duality of canonical
coordinates and momenta, i.e. the angular variable and the integers,
this formulation can also be interpreted as coherent states over an
infinite periodic chain. For the construction we use the analogy
with our quantization over a finite periodic chain where the phase
space was $\mathbb{Z}_{M} \times \mathbb{Z}_{M}$. Properties of the
coherent states constructed in this way are studied and the coherent
states are shown to satisfy the resolution of unity.
\end{abstract}

\section{Introduction}

Coherent states belong to the most important tools in numerous
applications of quantum theory.  The problem of coherent states on
the circle was investigated by S. de Bi\`evre \cite{SDB}, and also
by J. A. Gonz\'alez and M. A. del Olmo \cite{dOG}, based on
canonical coherent states \cite{Per1}. The Weil-Brezin-Zak transform
of canonical coherent states on the real line was then the essential
ingredient leading to a set of coherent states on the circle labeled
by the variables of the cylinder $\mathbb{R} \times \mathbf{S^{1}}$.

A different approach was employed by C. J. Isham and J. R. Klauder
\cite{IaK}. They constructed coherent states on the circle by using
the Euclidean group $E(2)$, which is the semi-direct product of
groups $\mathbb{R}^{2}$ and $SO(2)$ (see also \cite{Ni}) However,
they observed that there does not exist an irreducible
representation of $E(2)$ such that the resolution of unity holds.
Therefore they considered only reducible representations and
extended the method to the case of the $n$-dimensional sphere.

It should be noted that a very detailed study of deformation
quantization on the cylinder as classical phase space was recently
given in \cite{GdOT}. Their arguments confirm that in quantum theory
the `quantum' phase space $\mathbb{Z} \times S^{1}$ is involved,
which is our starting point in this investigation.

\section{Position and momentum operators on the circle}

Let the configuration space be the unit circle. The position
variable can be identified with the angle, i.e. it takes real values
modulo $2 \pi$. Let  $x$  take  a value from the interval $ \langle
-\pi, \pi )$. In the Dirac formalism the {\em position operator} is
defined
$$
  \mop{Q}  = \int _{-\pi}^{\pi}  x  \mket{x}  \mbra{x} d x,
\mbox{  with  } \sks{x}{y}= \delta(x-y).
 $$
The  position operator  \op{Q}  has continuous spectrum $x \in
\langle -\pi, \pi) $  with the corresponding eigenvectors
$\{\mket{x}\}.$ An arbitrary quantum  state \ket{\psi} can be
expressed in the form
$$
 \mket{\psi} =  \int _{-\pi}^{\pi} \psi (x) \mket{x},
 \quad \mbox{where} \quad \psi (x)=\sks{x}{\psi}.$$
It is useful to expand $\psi (x)$ in the Fourier series
$$
 \psi(x) = \sum_{n=-\infty}^{\infty} a_n e^{inx}.$$
The function $\psi (x)$ is the wavefunction  and the position
operator acts on it $ \mop{Q}\psi (x)=x \psi (x).$ The momentum
eigenvectors are then defined via Fourier transform
 \begin{eqnarray}
\mket{p}= \int_{-\pi}^{\pi} e^{ipx} \mket{x} d x
  \label{k:j}
\end{eqnarray}
and its inverse
   \begin{eqnarray}
\mket{x} =  \frac{1}{2\pi} \sum_{k=-\infty}^{\infty} e^{ikx}
\mket{k}.
         \label{j:k}
\end{eqnarray}

 In the Hilbert space formulation, where
 $$\mathcal{H}= L^2 (\mathbf S^1, d\varphi),$$
we shall follow our approach in \cite{TCh1}. The momentum operator
will be defined using the unitary representation $\mop{V}(\alpha)$
of the group of rotations $U(1)$ of the unit circle,
\begin{equation}\label{VVV}
  [\mathbf{V}(\alpha)\psi](\beta) = \psi(\beta-\alpha), \quad \psi \in
  L^{2}(\mathbf{S}^{1},d\varphi), \quad \alpha,\beta \in U(1).
\end{equation}
Unitary operators $\mathbf{V}(\alpha)$ shift the argument of
functions in $L^{2}(\mathbf{S}^{1},d\varphi)$. The momentum operator
is then by Stone's theorem
\begin{equation}\label{}
  \widehat{P} = -i\frac{d}{d\varphi}.
\end{equation}
The position operator
\begin{equation}\label{QQQ}
  (\widehat{\mathbf{Q}}\psi)(\varphi) = \varphi\psi(\varphi),
\end{equation}
formally satisfies the well known commutation relation
\begin{equation}\label{}
  [\widehat{Q},\widehat{P}]= i\mathbb{I},
\end{equation}
but it is not well-defined on $\mathcal{H}$.

\section{Construction of coherent states}

In order to define coherent states directly by Perelomov's method
\cite{Per1}, we should first construct a system of unitary operators
labeled by elements of the group $\mathbb{Z} \times U(1)$; second,
it is necessary to determine the `vacuum' vector $|0,0\rangle$. The
system of unitary operators will be defined
\begin{equation}\label{}
  \widehat{W}(m,\alpha) := e^{im\widehat{Q}}e^{-i\alpha
  \widehat{P}} = e^{im\widehat{Q}}\mathbf{V}(\alpha), \quad \alpha
  \in [\pi,\pi), \; m \in \mathbb{Z}.
\end{equation}
The factors do not commute
\begin{equation}\label{zyx}
 e^{im\widehat{\mathbf{Q}}}e^{-i\alpha\widehat{\mathbf{P}}}=
 e^{im\alpha}e^{-i\alpha\widehat{\mathbf{P}}}e^{im\widehat{\mathbf{Q}}},
 \quad \alpha \in [\pi,\pi),\; m \in \mathbb{Z},
\end{equation}
but the operator $e^{im\widehat{Q}}$ is now well defined on
$\mathcal{H}$,
\begin{equation}\label{expimq}
  e^{im\widehat{Q}}\psi(\varphi) = e^{im\varphi}\psi(\varphi).
\end{equation}

Here, the system of operators $\widehat{W}(m,\alpha)$ does not
create a representation of group $\mathbb{Z} \times U(1)$ but thanks
to the Weyl form of commutation relations we obtain a projective
representation of $\mathbb{Z} \times U(1)$.

The vacuum vector $|0,0\rangle$ will be determined in analogy with
canonical coherent states on $L^{2}(\mathbb{R})$. The requirement
that the vacuum state be an eigenvector of annihilation operator
with eigenvalue $0$ is written here in exponential form
\begin{equation}\label{qaip}
   e^{\widehat{Q} + i\widehat{P}}|0,0\rangle = |0,0\rangle.
\end{equation}
Using the Baker-Campbell-Hausdorff formula  the operator
$e^{\widehat{Q} + i\widehat{P}}$ could be separated in the product
of two operators $e^{\widehat{Q}}$ and $e^{i\widehat{P}}$ in
arbitrary order. Such change will not influence the final vacuum
state $|0,0\rangle$.

Condition (\ref{qaip}) leads to the Gauss exponential function
\begin{equation}\label{vacuum}
 |0,0\rangle = \mathcal{A}e^{-\frac{\varphi^{2}}{2}},
 \quad \varphi \in [-\pi , \pi).
\end{equation}
Hence the vacuum state is an element of our Hilbert space,
$|0,0\rangle \in L^{2}(\mathbf{S}^{1},d\varphi)$. At $\varphi = \pm
\pi$ it is continuous but its derivative has a small discontinuity
($\approx e^{-5}$). The normalization constant $\mathcal{A}$ is
given by
\begin{equation}\label{AaA}
  \mathcal{A}=\frac{1}{\sqrt{\int_{-\pi}^{\pi}exp(-\varphi^{2})d\varphi}}
  \doteq 0.751128.
\end{equation}

The system of coherent states on $L^{2}(\mathbf{S}^{1},d\varphi)$ is
now obtained by the action of the system of operators
$\widehat{W}(m,\alpha)$ on the vacuum state $|0,0\rangle$:
\begin{equation}\label{}
  |m,\alpha \rangle := \widehat{W}(m,\alpha) |0,0 \rangle.
\end{equation}
The functional form of our coherent states is given by
\begin{equation}\label{coherent}
  |m,\alpha \rangle =
  \mathcal{A}e^{im\varphi}e^{-\frac{(\varphi-\alpha)^{2}}{2}},
  \quad \varphi \in [- \pi,\pi ),
\end{equation}
i.e. for $\alpha \neq 0$ they are displaced and phased versions of
(\ref{vacuum}) with a discontinuity at $\varphi = \pm \pi$.

\section{Properties of our coherent states on
           $L^{2}(\mathbf{S}^{1},d\varphi)$}

In this section we shall check several properties of our coherent
states which are known to hold for canonical coherent states on
$L^{2}(\mathbb{R})$ \cite{Per1,Kla}. First of all, we shall look at
the resolution of unity, i.e. we shall prove the following equality
for our coherent states:
\begin{equation}\label{rou}
 \sum_{k \in \mathbb{Z}} \int_{\mathbf{S}^{1}}|k,\alpha \rangle
  \langle k,\alpha |d\alpha = c\widehat{I},
\end{equation}
where $c$ is a constant. Thus for an arbitrary normalized vector
$\eta$ from our Hilbert space $L^{2}(\mathbf{S}^{1},d\varphi)$ the
identity
\begin{equation}\label{roueta}
 \sum_{k \in \mathbb{Z}} \int_{\mathbf{S}^{1}}
 \langle \eta |k,\alpha \rangle
  \langle k,\alpha |\eta \rangle   d\alpha = c\widehat{I},
\end{equation}
should hold. Since the inner product of $|\eta\rangle$ with some
coherent state $|k,\alpha\rangle$ has the integral form
\begin{equation}\label{fouc}
   \langle k,\alpha | \eta \rangle =
  \mathcal{A}\int_{\mathbf{S}^{1}}e^{-ik\varphi}
  e^{-\frac{(\varphi - \alpha)^{2}}{2}}\eta(\varphi)d\varphi,
\end{equation}
one can calculate the constant $c$ with the result
\begin{equation}\label{}
  c = 2\pi,
\end{equation}
so that the resolution of unity for our set of coherent states takes
the form
\begin{equation}\label{dfg}
  \sum_{k \in \mathbb{Z}} \int_{\mathbf{S}^{1}}|k,\alpha \rangle
  \langle k,\alpha |d\alpha =
  2\pi \widehat{I}.
\end{equation}

The next properties to be checked concern the overlaps of our
(normalized) coherent states given by inner products in
$L^{2}(\mathbf{S}^{1},d\varphi)$. Here it is necessary to correctly
realize the way how the operator $\mathbf{V}(\alpha) =
exp(-i\alpha\widehat{P})$ acts on the Hilbert space
$L^{2}(\mathbf{S}^{1},d\varphi)$ when the circle $\mathbf{S^{1}}$
--- our configuration space --- is identified with the interval
$[-\pi,\pi)$. Then the action of operator $e^{-i\alpha\widehat{P}}$
on function $\psi(\varphi) \in L^{2}(\mathbf{S}^{1},d\varphi)$  for
 $\alpha \in [0,\pi]$ has the form :
\begin{eqnarray}\label{pro1}
  e^{-i\alpha\widehat{P}}\psi(\varphi) & = &
  \begin{cases}
    \psi(\varphi - \alpha) & \varphi \in [-\pi + \alpha,\pi], \\
    \psi(\varphi - \alpha +2\pi) & \varphi \in [-\pi , -\pi +
    \alpha].
  \end{cases}
\end{eqnarray}
For $\alpha \in [-\pi,0]$ we have
\begin{eqnarray}\label{pro2}
  e^{-i\alpha\widehat{P}}\psi(\varphi) & = &
  \begin{cases}
    \psi(\varphi - \alpha) & \varphi \in [-\pi ,\pi + \alpha], \\
    \psi(\varphi - \alpha -2\pi) & \varphi \in [\pi + \alpha , \pi].
  \end{cases}
\end{eqnarray}
Note that one has to consider addition modulo $2\pi$ in the argument
of function $\psi$. This is the reason why the inner product
\textit{cannot} be calculated simply according to
\begin{equation}\label{}
  \langle m,\alpha| n,\beta \rangle =
  \mathcal{A}^{2}\int_{-\pi}^{\pi}e^{-i\varphi(n-m)}e^{-\frac{(\varphi-\alpha)^{2}}{2}}e^{-\frac{(\varphi -
  \beta)^{2}}{2}}d\varphi.
\end{equation}

From now on we shall restrict ourselves only to the cases when
$\alpha$ and $\beta$ are non-negative:
\begin{equation}\label{230}
  \alpha \in [0, \pi], \quad \beta \in [0,\pi].
\end{equation}
Without loss of generality we may also assume
\begin{equation}\label{231}
  \beta \geq \alpha.
\end{equation}
Taking into account (\ref{pro1}) and (\ref{pro2}), we have to split
the inner product of two coherent states into two terms
\begin{equation}\label{232}
   \langle m,\alpha| n,\beta\rangle = \mathcal{A}I_{1}(\alpha,\beta,n-m) +
  \mathcal{A}I_{2}(\alpha,\beta,n-m),
\end{equation}
where
\begin{equation}\label{I1}
  I_{1}(\alpha,\beta,n-m) := \int_{\alpha - \pi}^{\beta -
  \pi}e^{i\varphi(n-m)} e^{-\frac{(\varphi - \alpha)^{2}}{2}}
  e^{-\frac{(\varphi-\beta + 2\pi)^{2}}{2}}d\varphi
\end{equation}
and
\begin{equation}\label{}
   I_{2}(\alpha,\beta,n-m) := \int_{\beta - \pi}^{\pi +
  \alpha}e^{i\varphi(n-m)} e^{-\frac{(\varphi-\alpha)^{2}}{2}}
   e^{-\frac{(\varphi-\beta)^{2}}{2}} d\varphi.
\end{equation}
In order to evaluate the integrals $I_{1}(\alpha,\beta,n-m)$ and
$I_{2}(\alpha,\beta,n-m)$, we use the error function of a complex
variable $z$,
\begin{equation}\label{erf}
  \mbox{erf}(z) := \frac{2}{\sqrt{\pi}}\int_{\Gamma
  (z)}e^{-\psi^{2}}d\psi;
\end{equation}
here $\Gamma(z)$ denotes an arbitrary continuous path of finite
length which connects the origin $0 \in \mathbb{C}$ with a complex
number $z \in \mathbb{C}$.

The integral $I_{1}(\alpha,\beta,n-m)$, after the substitution
\begin{equation}\label{}
  \omega = \varphi + \pi - \frac{\alpha + \beta}{2},
\end{equation}
takes the form
\begin{eqnarray}\label{}
  \nonumber I_{1}(\alpha,\beta,n-m)
 &= &
 e^{-(\frac{\beta-\alpha}{2})^{2}-\pi}e^{(\frac{\alpha+\beta}{2}-\pi)(m-n)}
 \times \\ & & \times
 \int_{\frac{\alpha-\beta}{2}}^{\frac{\beta-\alpha}{2}}e^{i\omega(n-m)}
 e^{-\omega^{2}}d\omega
\end{eqnarray}
which leads to the formula
\begin{eqnarray}\label{II11}
  \nonumber I_{1}(\alpha,\beta,n-m) & = &
(-\frac{\sqrt{\pi}}{2})e^{-(\frac{\beta-\alpha}{2})^{2}-\pi}
e^{i(\frac{\alpha+\beta}{2}-\pi)(m-n)}e^{-\frac{(n-m)^{2}}{4}}\times
\\ & &
\times[\mbox{erf}(\frac{\alpha-\beta}{2}+\frac{i(n-m)}{2})\\
& &+ \mbox{erf}(\frac{\alpha-\beta}{2}-\frac{i(n-m)}{2})].\nonumber
\end{eqnarray}
The other integral $I_{2}(\alpha,\beta,n-m)$, after the substitution
$  \omega = \varphi-\frac{\alpha+\beta}{2}$, yields
\begin{eqnarray}\label{II22}
  \nonumber I_{2}(\alpha,\beta,n-m) & = &
(-\frac{\sqrt{\pi}}{2})e^{-(\frac{\beta-\alpha}{2})^{2}}
e^{i(\frac{\alpha+\beta}{2})(m-n))}e^{-\frac{(n-m)^{2}}{4}}\times \\
& & \times[\mbox{erf}(\frac{\alpha-\beta}{2}-\pi+\frac{i(n-m)}{2})\\
& &
+\mbox{erf}(\frac{\alpha-\beta}{2}-\pi-\frac{i(n-m)}{2})].\nonumber
\end{eqnarray}

We have to admit that, unfortunately, we do not see any way how to
further simplify the above analytic expressions of the integrals
$I_{1}(\alpha,\beta,n-m)$ and $I_{2}(\alpha,\beta,n-m)$ to see
whether the coherent states are mutually non-orthogonal. However, we
have numerically computed absolute values of the inner products for
many pairs of coherent states and have drawn the graphs for several
different values of $n-m$ fixed in each graph. It was apparent
--- for all plotted cases --- that the overlap never vanishes.
Because of restriction on the length of the contribution we intend
to publish a detailed account of the matter including a number of
plots confirming this property.

The next checked quantities are the expectation values of the
position operator in the coherent states $|m,\alpha\rangle$ for
$\alpha \geq 0$. Explicitly we have
\begin{equation}\label{244}
  \langle m,\alpha|\widehat{Q}|m,\alpha\rangle = \mathcal{A}^{2}\int_{-\pi}^{-\pi +
  \alpha}\varphi e^{-(\varphi-\alpha + 2\pi)^{2}}d\varphi + \mathcal{A}^{2}\int_{-\pi +
  \alpha}^{\pi}\varphi e^{-(\varphi -\alpha)^{2}}d\varphi,
\end{equation}
or, using the error function,
\begin{equation}\label{245}
  \langle m,\alpha|\widehat{Q}|m,\alpha\rangle =
 \alpha- \mathcal{A}^{2}\sqrt{\pi^{3}}(\mbox{erf}(\pi)-\mbox{erf}(\pi -
 \alpha)).
\end{equation}
One observes that for $\alpha$ positive the expectation values do
not depend on $m$ (the same result is obtained also for negative
values of $\alpha$). It is interesting that the expectation value of
position is nonlinear in $\alpha$ for $\alpha \neq 0$. It is clear
that for $\alpha =0$ the quasi-Gaussian (\ref{vacuum}) is symmetric
around $\varphi = 0$, so the expectation value is an integral of an
odd function, evidently equal to zero. However, if $\alpha \in
(0,\pi]$, the displaced quasi-Gaussian is no more symmetric around
$\varphi = 0$ and the integration leads to a deviation depending on
a difference of error functions (\ref{245}); maximal deviation is
attained for $\alpha = \pi$.

Important among the checked quantities are the expectation values of
the momentum operator in coherent states $|m,\alpha\rangle$. The
explicit form
\begin{eqnarray}\label{}
  \nonumber \langle m,\alpha|\widehat{P}|m,\alpha\rangle &=& \mathcal{A}^{2}\int_{-\pi}^{-\pi +
  \alpha}(m +i(\varphi - \alpha + 2\pi)) e^{-(\varphi-\alpha + 2\pi)^{2}}d\varphi
  \\ \nonumber
   &+& \mathcal{A}^{2}\int_{-\pi +
  \alpha}^{\pi}(m+i(\varphi-\alpha)) e^{-(\varphi
  -\alpha)^{2}}d\varphi
\end{eqnarray}
can be simplified into the formula
\begin{equation}\label{}
  \langle m,\alpha|\widehat{P}|m,\alpha\rangle = m,
\end{equation}
which could be anticipated by analogy with the canonical coherent
states on $L^{2}(\mathbb{R})$.

Finally, the expectation values of the square of momentum operator
in coherent states $|m,\alpha\rangle$ are determined by computing
the integrals
\begin{eqnarray}\label{}
 \nonumber \langle m,\alpha|\widehat{P}^{2}|m,\alpha\rangle &=&
\mathcal{A}^{2}\int_{-\pi}^{-\pi + \alpha}[1+(m +i(\varphi -
\alpha + 2\pi))^{2}] e^{-(\varphi-\alpha + 2\pi)^{2}}d\varphi  \\
&+& \mathcal{A}^{2}\int_{-\pi +
  \alpha}^{\pi}[1+(m+i(\varphi-\alpha))^{2}] e^{-(\varphi
  -\alpha)^{2}}d\varphi
\end{eqnarray}
with the result
\begin{equation}\label{}
  \langle m,\alpha|\widehat{P}^{2}|m,\alpha\rangle =
 m^{2}+\frac{1}{2} +\mathcal{A}^{2}\pi e^{-\pi^{2}}
\end{equation}
which does not depend explicitly on $\alpha$.

\section{Conclusion}

For our system of coherent states their overlaps and matrix elements
were expressed analytically, then some calculated numerically or
evaluated with the help of MATHEMATICA. The property of resolution
of unity was shown to hold. \hsp

\section*{Acknowledgements} The support by the Ministry of Education
of Czech Republic (projects MSM6840770039 and LC06002) is
acknowledged. The authors are grateful to the referee for several
constructive remarks which helped to improve the presentation.

\end{document}